\documentclass{amsart}
\usepackage[utf8]{inputenc}
\usepackage[T1]{fontenc}

\usepackage{graphicx}
\graphicspath{{./Figures/}}

\usepackage[braket, qm]{qcircuit}
\usepackage{tikz-network}
\usepackage{pgfplots}
\usepackage{diagbox}
\usepackage{caption}
\usepackage{subcaption}
\usepackage{cleveref}
\usepackage{longtable,multirow,mathtools}
\usepackage[foot]{amsaddr}

\newtheorem{Definition}{Definition}

\newtheorem{Example}{Example}
\newtheorem{theorem}{Theorem}

\newtheorem{conjecture}{Conjecture}

\begin{document}

\title[Implementation of quantum hitting times on Cublike graphs]{Implementation of  hitting times of  discrete time quantum random walks on  Cubelike graphs.}

\author{Jaideep Mulherkar}
\email{jaideep\_mulherkar@daiict.ac.in}
\author{Rishikant Rajdeepak}
\email{rishikant\_rajdeepak@daiict.ac.in}
\author{V Sunitha}
\email{v\_suni@daiict.ac.in}
\address{Dhirubhai Ambani Institute of Information and Communication Technology , Gandhinagar, India.}

\maketitle

\begin{abstract}
We demonstrate an implementation of  the hitting time of a discrete time quantum random walk on  cubelike graphs using IBM's Qiskit platform. Our implementation is based on efficient circuits for the Grover and Shift operators. We verify the results about the one-shot hitting time of quantum walks on a hypercube as proved in \cite{Kempe2005}.  We extend the study to another family of cubelike graphs called the augmented cubes \cite{augmented_cubes}. Based on our numerical study, we conjecture that for all families of cubelike graphs there is a linear relationship between the degree of a cubelike graph and its hitting time which holds asymptotically. That is, for any  cubelike graph of degree $\Delta$, the probability of finding the quantum random walk at the target node at time $\frac{\pi \Delta}{2}$ approaches 1 as the degree $\Delta$ of the cubelike graph approaches infinity. 
\end{abstract}


%%%%%%%%%%%%%%%%%%%%%%%%%%%%%%%%%%%%%%%%%%%%%%%%%%%%%%%%%%%%%%%%%%%%%%%%%%%%%%%%%%%%%%%%%%%

\section{Introduction}
\label{sec:intro}
 Classical random walks have found applications in many areas such as computer science, physics, finance and engineering. In particular, classical random walks have been used in the development of faster algorithms for estimating the volume of a convex body, finding a satisfying assignment to a Boolean formula (2-SAT problem) and finding if two vertices of an undirected graph are connected \cite{Lovasz}. Quantum random walks  \cite{Kempe2003,Renato,Andraca_2012} are quantum mechanical analogues of classical random walks. The first quantum random walk models were proposed  in \cite{ADZ1993,FG1998}. It has since been observed that there are some startling differences between classical and quantum walks. For instance, it has been shown that the hitting time from one vertex to the antipodal vertex is linear in the dimension of the hypercube, which is exponentially fast in quantum walks compared to classical walks \cite{Kempe2005}. Moreover, there is also a speed up obtained in the mixing time of quantum versus the classical walks \cite{Moore}. These differences between classical and quantum walks led to search for quantum walk based algorithms that can outperform classical counterparts. Some of the quantum algorithms that have been developed based on quantum walks are searching a marked element  on a grid, the element distinctness problem and the triangle finding problem in a graph \cite{AA2003}. Quantum walks have also been shown to be universal for quantum computation \cite{AC2009,NCETV2010}.

 In  quantum random walks there is a quantum system such as a spin system on a graph and the evolution is given by a unitary operator acting on a Hilbert space of the graph. Generally speaking quantum walks are of two types : Discrete and Continuous. In the discrete time quantum walk the associated Hilbert space is a joint system of a coin space and position space.  For a regular graph the coin space has dimension equal to the number of edges per node and the dimension of the position space is equal to the number of vertices in the graph.  The discrete time quantum walk starts with an initial state of the coin and  position. Each time step in the evolution of the discrete random walk consists of the following: 1. A coin operator acts on the coin space changing the original coin state to a new state 2. A Shift operator shifts the current position of the random walker conditioned on the coin state. This step is then repeated many times thus evolving the random walker  over the state space of the graph. After a certain number of time steps a measurement is done of the position space. The continuous random walks are the evolution of a quantum system on a graph according to the unitary $e^{iAt}$ where $A$ is the adjacency matrix of the graph. A quantity of interest in quantum walks is the hitting time for a pair of vertices. For a pair of vertices $(u,v)$, if starting from vertex $u$,  after a certain number of time steps $T$ of the walker is observed to be in vertex $v$ with probability $p$ then $(T,p)$ is called the hitting time for vertex $v$ starting from vertex $u$. The hitting times are based on the graph topology and faster hitting times could mean speed up algorithms.  Quantum walks have been studied on a variety of graphs such as cycles, trees, and cubelike graphs and general Cayley graphs \cite{Aharonov_2001,Acevedo_2005}. A related question  is  of perfect state transfer between two vertices, which has been studied in \cite{CG2012}. Apart from the results in \cite{Kempe2005} in hypercubes, cubelike graphs have been studied in the context of continuous time random walks. Criterion for the perfect state transfer on cubelike graphs have been given in \cite{BGS2008,CheungGodsil_2011}. In this paper we restrict our study to the hitting times of discrete time quantum  walks on cubelike graphs.

 Because of their universality and potential use in developing algorithms there is a need for good physical implementations of quantum random walks.  Physical implementations based on ion-traps, optical cavities and optical lattices has been suggested in \cite{KRS2003,RS2005,XSL2009}. Circuit implementations of various standard quantum algorithms using IBM's Qiskit platform was done in \cite{Abhijith2020quantum}. Recently in \cite{AAKP2020} implementation of staggered quantum walks  on cycles, two-dimensional lattices and complete graph was studied on IBM quantum computers. A comparison on two different implementation approaches of quantum walks is given in \cite{GEZ2021}. An implementation of the discrete time quantum walk on hypercubes, complete graphs, complete bipartite graphs and 2-d lattice is recently given in \cite{Wanzambi_2021}. Our work is in the spirit of implementation of QRW's for NISQ (Nosiy Intermediate Scale Quantum computing) era computers. Our main contributions in this paper are  1) An implementation of efficient circuits for the QRW on a cubelike graph based on decompositions of the Grover and Shift operator. 2) An application of these circuits to find the hitting times in cubelike graphs. Here we implement our circuits on the IBM's Qiskit platform and verify the one-shot hitting times in families of cubelike graphs such as hypercubes and augmented cubes.

This paper is organized as follows: In section \ref{sec:Setup} we describe the  construction of the discrete time quantum random walk based on the Grover coin operator and the Shift operator on a cubelike graph. In section \ref{sec:Hitting_times} we define the one-shot hitting time of quantum random walks and review the past results. In section \ref{sec:QCircuits} we describe our implementation of the efficient circuits for the QRW and finally in section \ref{sec:Results} we apply our efficient circuits to the problem of hitting times for various cubeike graphs.

%%%%%%%%%%%%%%%%%%%%%%%%%%%%%%%%%%%%%%%%%%%%%%%%%%%%%%%%%%%%%%%%%%%%%%%%%%%%%%%%%%%%%%%%%%%
%%%%%%%%%%%%%%%%%%%%%%%%%%%%%%%%%%%%%%%%%%%%%%%%%%%%%%%%%%%%%%%%%%%%%%%%%%%%%%%%%%%%%%%%%%%

\section{Discrete-time Quantum Walk on Cubelike graphs}
\label{sec:Setup}

\subsection{The quantum walk} \label{subsec:DTQW} A discrete-time quantum walk (DTQW) on a graph $\Gamma$ is described by the evolution of an associated quantum system in a Hilbert space. If $\Gamma$ is $\Delta$-regular with $N$ vertices, then the associated Hilbert space is the joint Hilbert space $\mathcal{H}=\mathcal{H}_C\otimes \mathcal{H}_P$, where $\mathcal{H}_C\cong \mathbb{C}^{\Delta}$ is known as the coin space, and $\mathcal{H}_P\cong \mathbb{C}^{N}$ is known as the position space. Suppose vertices and edges in $\Gamma$ are labeled by $\{v_0,v_1,\dots,v_{N-1}\}$ and $\{\alpha_0,\alpha_1,\dots,\alpha_{\Delta-1}\}$, respectively, then the computational basis of $\mathcal{H}$ is given by
\begin{equation}
\label{eq:basis}
    \{\ket{\alpha_k}\ket{v_a}:0\leq k \leq \Delta-1,\;0\leq a\leq N-1\}.
\end{equation}
This association between $\Gamma$ and $\mathcal{H}$ is such that vertex $v_a$ represents the position state $\ket{v_a}$ in $\mathcal{H}_P$, and edge $\alpha_k$ represents the coin state $\ket{\alpha_k}$ in $\mathcal{H}_C$. The evolution of the system is described by a unitary operator $U$, called the evolution operator, defined by
\begin{equation}
    \label{eq:evolution_operator}
    U=S(C\otimes I),
\end{equation}
where, $C$ is the coin operator analogous to $\Delta$-dimensional classical coin and $S$ is the shift operator that shifts the quantum state $\ket{\alpha_k}\ket{v_a}$ to the quantum state $\ket{\alpha_k}\ket{v_b}$ such that $\alpha_k$ is the label of the edge $(v_a,v_b)$.

% A generic quantum state of the quantum system associated with the graph is the linear combination of the basis vectors, i.e.,
% \begin{equation}
% \label{eq:generic_state}
%     \ket{\Psi}=\sum_{e=0}^{d-1}\sum_{v=0}^{N-1}\lambda_{e,v}\ket{e}\ket{v},\qquad\lambda_{e,v}\in\mathbb{C}.
% \end{equation}

% In this report, we use the classical terminology to mean the quantum counterparts such as (a) the  walker for the quantum system, (b) the vertex for the position state, (c) the direction or edge for the coin state, (d) the walk for the quantum evolution, and so on.

\begin{figure}[t]
\tiny
\centering
    \begin{subfigure}[t]{.45\textwidth}
    \begin{tikzpicture}[scale=.7]
        \begin{scope}[every node/.style = {draw, circle, inner sep=1pt}]
        \node (0) at (-1,1) [label=left:$000$]{$v_0$};
        \node (1) at (1,1) [label=right:$001$]{$v_1$};
        \node (3) at (1,-1) [label=right:$011$]{$v_3$};
        \node (2) at (-1,-1) [label=left:$010$]{$v_2$};
        \node (4) at (-3,3) [label=left:$100$]{$v_4$};
        \node (5) at (3,3) [label=right:$101$]{$v_5$};
        \node (7) at (3,-3) [label=right:$111$]{$v_7$};
        \node (6) at (-3,-3) [label=left:$110$]{$v_6$};
    \end{scope}
        \draw (0) edge node [above] {$\alpha_0$} (1); \draw (0) edge node [below] {$00$} (1); 
        \draw (0) edge node [left] {$\alpha_1$} (2); \draw (0) edge node [right] {$01$} (2);
        \draw (0) edge node [above right] {$\alpha_2$} (4); \draw (0) edge node [below left] {$10$} (4); 
        \draw (1) edge node [right] {$\alpha_1$} (3); \draw (1) edge node [left] {$01$} (3);
        \draw (1) edge node [above left] {$\alpha_2$} (5); \draw (1) edge node [below right] {$10$} (5);
        \draw (2) edge node [below] {$\alpha_0$} (3); \draw (2) edge node [above] {$00$} (3);
        \draw (2) edge node [above left] {$\alpha_2$} (6); \draw (2) edge node [below right] {$10$} (6); 
        \draw (3) edge node [above right] {$\alpha_2$} (7); \draw (3) edge node [below left] {$10$} (7);
        \draw (4) edge node [above] {$\alpha_0$} (5); \draw (4) edge node [below] {$00$} (5); 
        \draw (4) edge node [left] {$\alpha_1$} (6); \draw (4) edge node [right] {$01$} (6);
        \draw (5) edge node [right] {$\alpha_1$} (7); \draw (5) edge node [left] {$01$} (7); 
        \draw (6) edge node [below] {$\alpha_0$} (7); \draw (6) edge node [above] {$00$} (7);
    \end{tikzpicture}
    \caption{\label{fig:Q3} $\mathcal{Q}_3$}
    \end{subfigure}
    \begin{subfigure}[t]{.45\textwidth}
    \begin{tikzpicture}[scale=.7]
        \begin{scope}[every node/.style = {draw, circle, inner sep=1pt}]
        \node (0) at (-1,1) [label=above:]{$v_0$};
        \node (1) at (1,1) [label=above:]{$v_1$};
        \node (3) at (1,-1) [label=below:]{$v_3$};
        \node (2) at (-1,-1) [label=below:]{$v_2$};
        \node (4) at (-3,3) [label=above:]{$v_4$};
        \node (5) at (3,3) [label=above:]{$v_5$};
        \node (7) at (3,-3) [label=below:]{$v_7$};
        \node (6) at (-3,-3) [label=below:]{$v_6$};
    \end{scope}
        \draw (0) edge node [above] {$\alpha_0$} (1); 
        \draw (0) edge node [left] {$\alpha_1$} (2); 
        \draw (0) edge node [above right] {$\alpha_3$} (4);
        \draw[red] (0) edge[bend left=20] node [below] {$\alpha_2$} (3);
        \draw (0) edge[red,bend right=20] node [below] {$\alpha_4$} (7);
        \draw (1) edge node [right] {$\alpha_1$} (3); 
        \draw (1) edge node [above left] {$\alpha_3$} (5);
        \draw[red] (1) edge[bend right=20]  node [below] {$\alpha_2$} (2);
        \draw[red] (1) edge [bend left=20] node [below] {$\alpha_4$} (6);
        \draw (2) edge node [below] {$\alpha_0$} (3);
        \draw (2) edge node [above left] {$\alpha_3$} (6);
        \draw (2) edge[red,bend right=20] node[right] {$\alpha_4$} (5);
        \draw (3) edge[red,bend left=20] node[below] {$\alpha_4$} (4);
        \draw (3) edge node [above right] {$\alpha_3$} (7); 
        \draw (4) edge node [above] {$\alpha_0$} (5);
        \draw (4) edge node [left] {$\alpha_1$} (6);
        \draw (4) edge[red,bend right=45] node[below] {$\alpha_2$} (7);
        \draw (5) edge node [right] {$\alpha_1$} (7); 
        \draw (5) edge[red, bend left=45] node [below] {$\alpha_2$} (6);
        \draw (6) edge node [below] {$\alpha_0$} (7); 
    \end{tikzpicture}
    \caption{\label{fig:AQ3} $\mathcal{AQ}_3$}
    \end{subfigure}
    \caption{Pictorial representation of (\subref{fig:Q3}) $\mathcal{Q}_3=Cay(\mathbb{Z}_2^3,\{001,010,100\})$ and (\subref{fig:AQ3}) $\mathcal{AQ}_3=Cay(\mathbb{Z}_2^3,\{001,010,011,100,111\})$. In both the cases, a vertex $x_2x_1x_0$ is represented by $v_a$, where $a=x_22^2+x_12^1+x_02^0$, and an edge $(v_a,v_b)$ is represented by $\alpha_{k-1}$, if $v_a\oplus v_b$ is the $k$-element of the corresponding lexicographically ordered generating set. The edge $\alpha_{k-1}$ is also represented by the binary string (of length 2) of the integer $k-1$.}
    \label{fig:labeling}
\end{figure}
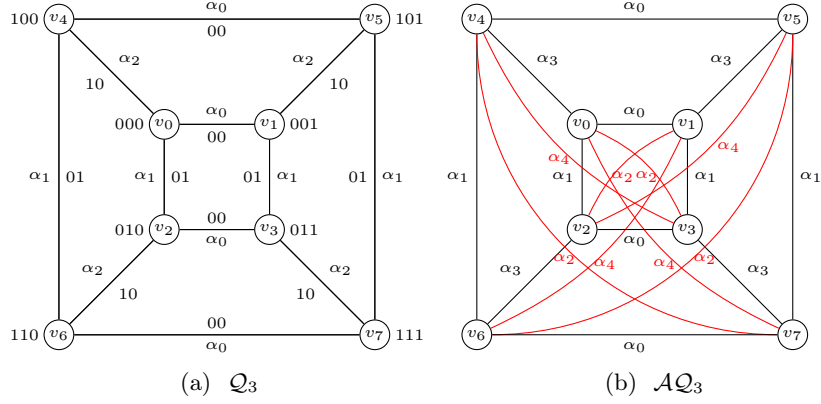

\subsection{Cubelike graphs}
\label{subsec:cubelike}
A Cayley graph is a graph defined over a pair $(G,\Omega)$, where $G$ is a finite group and $\Omega$ is a generating set of $G$, with the properties; (1) $\Omega$ does not contain the identity element, and (2) $\Omega$ is closed under the group inverse, i.e., $x^{-1}\in\Omega$ for all $x\in \Omega$. The Cayley graph, denoted by $Cay(G,\Omega)$, is a graph whose vertices are identified with the elements of $G$ and the edge set is $\{(x,y):xy^{-1}\in\Omega\}$. Clearly, $Cay(G,\Omega)$ is a regular graph on $N=|G|$ vertices with regularity $\Delta=|\Omega|$. The Cayley graph $Cay(\mathbb{Z}_2^n,\Omega)$ defined over the Boolean group $\mathbb{Z}_2^n$ is called cubelike graph of dimension $n$. For a cubelike graph $Cay(\mathbb{Z}_2^n,\Omega)$ when the ordering on $\Omega$ is not specified then we assume the lexicographical ordering, i.e., we assume that the cubelike graph is given in its canonical form. A vertex $(x_{n-1},\dots,x_1,x_0)\in\mathbb{Z}_2^n$ is denoted by the binary string $x_{n-1}\cdots x_1x_0$, and is given the label $v_a$, where $a=x_{n-1}2^{n-1}+\cdots + x_12^1 + x_02^0$. An edge $(v_a,v_b)$ is labeled by $\alpha_{k-1}$, $1\leq k \leq \Delta$, if the XOR operation of $v_a$ and $v_b$, i.e. $v_a\oplus v_b$, is the $k$-th element $\Omega(k)$ of $\Omega$. Suppose $2^{m-1}<\Delta\leq 2^m$, for some integer $m$, then the edge labeled by $\alpha_{k-1}$ can be denoted by the $k$-th binary string $y_{m-1}\cdots y_1y_0$ in $\mathbb{Z}_2^m$, i.e., $k-1=y_{m-1}2^{m-1}+\cdots+y_12^1+y_02^0$. 

Some special subfamily of cubelike graphs of interest are (a) Hypercubes $\mathcal{Q}_n$ with $\Omega=\{0^i10^j:i+j=n-1\}$, and (b) Augmented cubes~\cite{augmented_cubes} $\mathcal{AQ}_n$ with $\Omega=\{0^i10^j:i+j=n-1\}\cup\{0^{n-i}1^i:1\leq i \leq n\}$. 
In Figure~\ref{fig:Q3}, edge $(v_0,v_4)$ is labeled by $\alpha_2$ because $v_0\oplus v_4=100$ is the third element of its generating set $\{001,010,100\}$,  while in Figure~\ref{fig:AQ3} it is labeled by $\alpha_3$ as $100$ is the fourth element of its generating set $\{001,010,011,100,111\}$. The binary representations of $\alpha_0$, $\alpha_1$, $\alpha_2$ in $\mathcal{Q}_3$ (Figure~\ref{fig:Q3}) are $00$, $01$, $10$, respectively, while in $\mathcal{AQ}_3$ (Figure~\ref{fig:AQ3}), the binary representations of $\alpha_0$, $\alpha_1$, $\alpha_2$, $\alpha_3$, $\alpha_4$ are $000$, $001$, $010$, $011$, $100$, respectively.

%In this report, we use the classical terminology to mean the quantum counterparts such as (a) the  walker for the quantum system, (b) the vertex for the position state, (c) the direction or edge for the coin state, (d) the walk for the quantum evolution, and so on.

% The Hilbert space associated with $Cay(\mathbb{Z}_2^n,\Omega)$, with $d=|\Omega|$, has the computational basis
% \begin{equation}\{\ket{\alpha_k}\ket{v_i}: 0\leq k\leq d-1,\;0\leq i\leq 2^n-1\}.\end{equation}

Let $U=S(C\otimes I)$ be the evolution operator for DTWQ on $Cay(\mathbb{Z}_2^n,\Omega)$, with $\Delta=|\Omega|$. Then, the shift operator $S$ is defined by
\begin{equation}
    S=\sum_{k=1}^{\Delta}\sum_{a=0}^{2^n-1} \ket{\alpha_{k-1},v_a\oplus \Omega(k)}\bra{\alpha_{k-1},v_a},
    \label{eq:shift_operator}
\end{equation}
Any unitary operator on $\mathbb{C}^{\Delta}$ can be chosen as the coin operator. A
popular coin operator is the Grover coin $C$ which is defined by $2\ket{D}\bra{D}-I$, where,
\begin{equation}\label{eq:grover}
\ket{D}=\frac{1}{\sqrt{\Delta}}\sum_{k=1}^{\Delta}\ket{\alpha_{k-1}} \implies 
C= \begin{bmatrix} \frac{2}{\Delta}-1 & \frac{2}\Delta & \frac{2}\Delta & \cdots & \frac{2}\Delta \\[.5cm] \frac{2}\Delta & \frac{2}\Delta-1 & \frac{2}\Delta & \cdots & \frac{2}\Delta \\[.5cm]
\vdots & \vdots & \vdots & \ddots & \vdots \\[.5cm] \frac{2}\Delta & \frac{2}\Delta & \frac{2}\Delta & \cdots & \frac{2}\Delta-1\end{bmatrix}.
\end{equation}

 In the following example DTQW is illustrated on $C_4$, a cycle on four vertices.
\begin{Example}[Discrete-time quantum walk on $Cay(\mathbb{Z}_2^2,\{01,10\})$]
\label{eg:DTQW_Q2}
The Hilbert space associated with the graph is $\mathbb{C}^2\otimes \mathbb{C}^4$, and the computational basis is  \[\{\ket{0}\ket{00},\ket{1}\ket{00},\ket{0}\ket{01},\ket{1}\ket{01},\ket{0}\ket{10},\ket{1}\ket{10}, \ket{0}\ket{11},\ket{1}\ket{11}\}.\]
The evolution operator $U$ is $S(C\otimes I)$, where,
\[
C = \begin{bmatrix} 0 & 1 \\ 1 & 0  \end{bmatrix},
\]
and,
\[ \begin{split} S &= \sum_{k=1}^2\sum_{a=0}^3\ket{\alpha_{k-1}}\ket{v_a\oplus \Omega(k)}\bra{\alpha_{k-1}}\bra{v_a}.
\end{split}\]

% \\
% &=\ket{0}\ket{00\oplus 01}\bra{0}\bra{00} + \ket{0}\ket{01\oplus 01}\bra{0}\bra{01} \\ 
% &+ \ket{0}\ket{10\oplus 01}\bra{0}\bra{10} + \ket{0}\ket{11\oplus 01}\bra{0}\bra{11}
% \\
% &+\ket{1}\ket{00\oplus 10}\bra{1}\bra{00} + \ket{1}\ket{01\oplus 10}\bra{1}\bra{01} \\ 
% &+ \ket{1}\ket{10\oplus 10}\bra{1}\bra{10} + \ket{1}\ket{11\oplus 10}\bra{1}\bra{11}.

The evolution, with the initial state $\ket{D}\ket{00}=\frac{\ket{0}+\ket{1}}{\sqrt{2}}\ket{00}$, occurs as:
\begin{enumerate}
\itemsep1em
        \item The first time step; 
        %\begin{equation}
        \[\begin{split}
            U\ket{D}\ket{00} & = 
            S(C\otimes I)\left(\frac{1}{\sqrt{2}}(\ket{0}\ket{00}+\ket{1}\ket{00})\right) \\
            & =\frac{1}{\sqrt{2}}\left(\ket{1}\ket{10}+\ket{0}\ket{01}\right).
        \end{split}\]
        %\end{equation}
        \item The second time step;
        %\begin{equation}
        \[\begin{split} 
            U^2 \ket{D}\ket{00}
         	&= S \frac{1}{\sqrt{2}}\left[\left(2\frac{\ket{0}+\ket{1}}{\sqrt{2}}\frac{1}{\sqrt{2}} - \ket{1} \right)\ket{10}\right. \\
         	&+\left. \left(2\frac{\ket{0}+\ket{1}}{\sqrt{2}}\frac{1}{\sqrt{2}}-\ket{0} \right)\ket{01} \right] \\
        	&= S\frac{1}{\sqrt{2}}\left(\ket{0}\ket{10}+\ket{1}\ket{01}\right)
        	=\frac{\ket{0}\ket{11}+\ket{1}\ket{11}}{\sqrt{2}} \\ 
        	&=\ket{D}\ket{11}.
        \end{split}\]
        %\end{equation}
\end{enumerate}
Consequently, we get the probability distribution of the quantum states. After the first step, the quantum system is at state $\ket{01}$ or $\ket{10}$ with probability $\frac{1}{2}$, and after two steps the probability of the quantum system to be at $\ket{11}$ is $1$.
\end{Example}

\section{Hitting times of quantum walks}
\label{sec:Hitting_times}
The DTQW starts from a particular starting vertex $v_0$ and evolves under the dynamics of the evolution operator given in equations (\ref{eq:shift_operator}) and (\ref{eq:grover}). Letting the system evolve for a finite number of steps $T$ is equivalent to  applying $U^T$ to an initial state $\ket{v_0}$ of the system. A measurement is then performed in the position space basis to see if the walker is found in a specific target vertex. At time T, if we find the walker in the state  $\ket{v}$ corresponding to the target vertex $v$ with probability at least $p$, then we say $(T,p)$ is the quantum hitting time for $v$, starting from vertex $v_0$.  This hitting time was called the one-shot hitting time defined by Kempe in \cite{Kempe2005}. More formally,

\begin{Definition}[\cite{Kempe2005}]
 The discrete-time quantum walk is said to admit $(T,p)$ one-shot $(\ket{v_0},\ket{v})$ hitting time if $|\bra{v}U^T\ket{v_0}|^2\geq p$. 
 \end{Definition}
 It was shown in \cite{Kempe2005} that the one-shot hitting time a of quantum random walk on a hypercube is exponentially faster than its classical analogue. We state the following result from \cite{Kempe2005}.
 \begin{theorem}[\cite{Kempe2005}]
    The discrete-time quantum walk with Grover coin on the hypercube of dimension $n$ has a $(T, p)$ one-shot $(\ket{x},\ket{\bar{x}})$ hitting time, i.e. $\Big|\bra{\bar{x}}U^T\ket{x}\Big|^2\geq p$, where
    $T$ is either $\lfloor{\frac{\pi}{2}n}\rfloor$ or $\lceil{\frac{\pi}{2}n}\rceil$  and $p=1-\mathcal{O}(\frac{\log^3n}{n})$.
\end{theorem}

\begin{figure}[t]
\tiny
\begin{equation*}
    \Qcircuit @C=0.5em @R=0.2em @!R {
	 	\lstick{ {q}_{0} :  } & \qw \barrier[0em]{2} & \qw & \targ & \qw & \qw \barrier[0em]{2} & \qw & \targ & \qw & \qw \barrier[0em]{2} & \qw & \meter & \qw & \qw & \qw\\
	 	\lstick{ {q}_{1} :  } & \qw & \qw & \qw & \qw & \targ & \qw & \qw & \qw & \targ & \qw & \qw & \meter & \qw & \qw\\
	 	\lstick{ {q}_{2} :  } & \gate{\mathrm{H}} & \qw & \ctrl{-2} & \gate{\mathrm{X}} & \ctrl{-1} & \qw & \ctrl{-2} & \gate{\mathrm{X}} & \ctrl{-1} & \qw & \qw & \qw & \qw & \qw\\
	 	\lstick{c:} & \lstick{/_{_{2}}} \cw & \cw & \cw & \cw & \cw & \cw & \cw & \cw & \cw & \cw & \dstick{_{_{0}}} \cw \cwx[-3] & \dstick{_{_{1}}} \cw \cwx[-2] & \cw & \cw\\
	 }
\end{equation*}
\caption{The quantum circuit for DTWQ on $\mathcal{Q}_2$. }
\label{fig:Q2_circuit}
\end{figure}
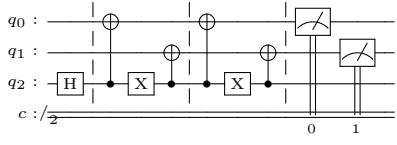

For example, $\mathcal{Q}_2$ has $(2,1)$ one-shot $(\ket{00},\ket{11})$ hitting time. The quantum circuit for the DTWQ on $\mathcal{Q}_2$ is shown in Figure~\ref{fig:Q2_circuit}~\cite{Abhijith2020quantum}. The first two qubits represent the position state which is initially at zero state $\ket{00}$. The third qubit represents the coin state, and is initialized to the diagonal state by applying the Hadamard gate $H$. The composition of the next three operators, namely, two $CNOT$-gates and one $NOT$-gate, forms the evolution operator $U$. The evolution operator $U$ is applied twice, which are separated by a dotted vertical line called the barrier. The following calculations show that the circuit (Figure~\ref{fig:Q2_circuit}) is indeed a DTQW on $\mathcal{Q}_2$. We use $X(a)$ to denote the X-gate that flips the value of the qubit $a$, and $C_X(a,b)$ to denote the CNOT-gate with control qubit $a$ and target qubit $b$.
\begin{enumerate}
	\item The system is initialized to $(H\otimes I\otimes I) \ket{000}$, where $I$ is the identity operator. Since,
		\[(H\otimes I\otimes I) \ket{000} = \frac{\ket{000}+\ket{100}}{\sqrt{2}} = \ket{D}\ket{00},\]
		the system is thus initialized to $\ket{D}\ket{00}$.
	\item The application of $U$ to this initial state evolves the system to the quantum state $\frac{\ket{110}+\ket{001}}{\sqrt{2}}$ as shown below.  
		\[
		\begin{split}
			\frac{\ket{000}+\ket{100}}{\sqrt{2}} &\xrightarrow{C_X(2,0)} \frac{\ket{000}+\ket{101}}{\sqrt{2}} \\
			    &\xrightarrow{X(2)} \frac{\ket{100}+\ket{001}}{\sqrt{2}} \\
				&\xrightarrow{C_X(2,1)} \frac{\ket{110}+\ket{001}}{\sqrt{2}},
		\end{split}
		\]
	\item By applying $U$ to the current state, we see that the system evolves to the state $\ket{D}\ket{11}$, as shown below.
	\[
	\begin{split}
		\frac{\ket{110}+\ket{001}}{\sqrt{2}} &\xrightarrow{C_X(2,0)} \frac{\ket{111}+\ket{001}}{\sqrt{2}} \\
		&\xrightarrow{X(2)} \frac{\ket{011}+\ket{101}}{\sqrt{2}} \\
		&\xrightarrow{C_X(2,1)} \frac{\ket{011}+\ket{111}}{\sqrt{2}} = \ket{D}\ket{11}.
	\end{split}
	\]
\end{enumerate}
Now, the measurement operator outputs $\ket{11}$ with probability $1$. If the measurement was done after the first application of $U$ and then after the second application, then, we would have obtained either $\ket{00}$ or $\ket{11}$ with probability $\frac{1}{2}$. 

%%%%%%%%%%%%%%%%%%%%%%%%%%%%%%%%%%%%%%%%%%%%%%%%%%%%%%%%%%%%%%%%%%%%%%%%%%%%%%%%%%%%%%%%%%%
%%%%%%%%%%%%%%%%%%%%%%%%%%%%%%%%%%%%%%%%%%%%%%%%%%%%%%%%%%%%%%%%%%%%%%%%%%%%%%%%%%%%%%%%%%%
\section{The Quantum circuit for DTQW on cubelike graphs}
\label{sec:QCircuits}
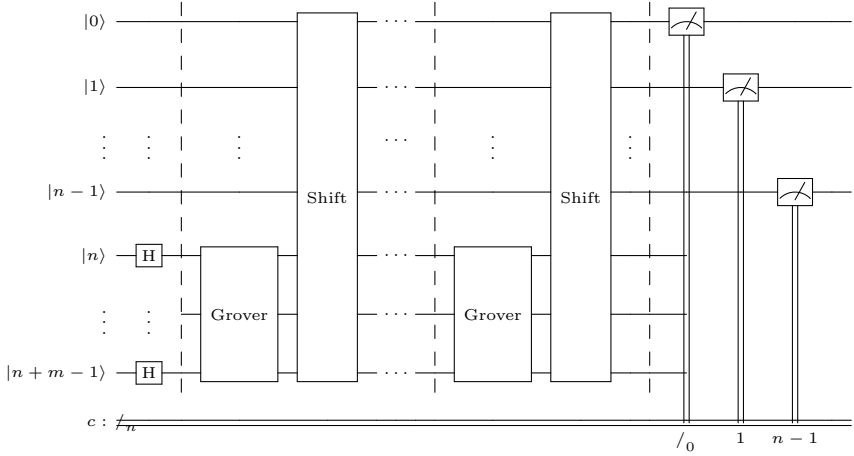
\begin{figure}[htp]
\tiny
\[
\Qcircuit @C=1.0em @R=2em
{ 
\lstick{\ket{0}} & \qw \barrier[0em]{6} & \qw & \qw & \multigate{6}{\mathrm{Shift}} & \qw & \cdots & \barrier[0em]{6} & \qw & \qw & \multigate{6}{\mathrm{Shift}} & \qw \barrier[0em]{6} & \qw & \meter & \qw & \qw & \qw & \qw \\
\lstick{\ket{1}} & \qw & \qw & \qw & \ghost{\mathrm{Shift}} &  \qw & \cdots & & \qw & \qw & \ghost{\mathrm{Shift}}  & \qw & \qw & \qw & \meter & \qw & \qw & \qw \\
\lstick{\vdots} & \vdots &  & \vdots & &  & \cdots & &  & \vdots & & \vdots &  & \\
\lstick{\ket{n-1}} & \qw & \qw  & \qw & \ghost{\mathrm{Shift}} &  \qw & \cdots & & \qw  & \qw & \ghost{\mathrm{Shift}} & \qw & \qw & \qw & \qw & \meter & \qw & \qw \\
\lstick{\ket{n}} & \gate{\mathrm{H}} & \qw & \multigate{2}{\mathrm{Grover}} & \ghost{\mathrm{Shift}} &  \qw & \cdots & & \qw & \multigate{2}{\mathrm{Grover}} & \ghost{\mathrm{Shift}} & \qw & \qw & \qw \\
\lstick{\vdots} & \vdots & & \ghost{\mathrm{Grover}} & \ghost{\mathrm{Shift}} &  \qw & \cdots & & \qw & \ghost{\mathrm{Grover}} & \ghost{\mathrm{Shift}} & \qw & \qw & \qw \\
\lstick{\ket{n+m-1}} & \gate{\mathrm{H}} & \qw  & \ghost{\mathrm{Grover}} & \ghost{\mathrm{Shift}} &  \qw & \cdots & & \qw  & \ghost{\mathrm{Grover}} & \ghost{\mathrm{Shift}} & \qw & \qw & \qw \\
\lstick{c:} & \lstick{/_{_n}} \cw & \cw & \cw & \cw & \cw & \cw & \cw & \cw & \cw & \cw & \cw & \cw & \dstick{/_{_0}} \cw \cwx[-7] & \dstick{1} \cw \cwx[-6] & \dstick{n-1} \cw \cwx[-4] & \cw & \cw \\
}
\]
\caption{Quantum circuit for DTQW on cubelike graph. H is applied to each coin qubit to initialize the coin state to $\ket{D}$. The evolution operator $U=S(C\otimes I)$ is applied for a finite number of times. Each application of U is separated by a barrier. Finally, measurement operators are applied to the position qubits.}
\label{fig: general_cubelike_circuit}
\end{figure}

A general quantum circuit representing the DTQW on $Cay(\mathbb{Z}_2^n,\Omega)$, with $\Delta=|\Omega|$, is depicted in Figure~\ref{fig: general_cubelike_circuit}. A unitary operator that can be appended to a quantum circuit has dimension $2^t$, for some positive integer $t$. Since the coin operator acting on the coin space is of dimension $\Delta$, we study the case where $\Delta=2^m$, for some positive integer $m$. All the qubits are  initialized to state $\ket{0}$. The position state is represented by the first $n$ qubits, which is initially $\ket{v_0}=\ket{0}^{\otimes n}$. The coin state, represented by the last $m$ qubits, is initialized to diagonal state  $\ket{D}=\frac{1}{\sqrt{\Delta}}\sum_{k=1}^{\Delta}\ket{\alpha_{k-1}}$ by applying the Hadamard gate H to each coin qubit. After the initialization, the evolution operator $U$, which is the composition of the Grover operator $C$ and the shift operator $S$, is applied to the circuit for a specific number of times before the measurement is taken. It is important to decompose the evolution operator to run the circuit more efficiently.

\subsection{Decomposition of the evolution operator $U$}
\subsubsection{Decomposition of the Grover operator}
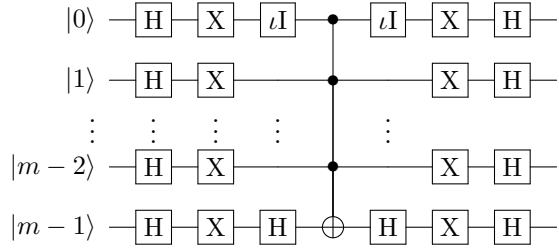
\begin{figure}[htp]
\[
\Qcircuit @C=1em @R=1em
{
\lstick{\ket{0}} & \gate{\mathrm{H}} & \gate{\mathrm{X}} & \gate{\iota\mathrm{I}} & \ctrl{4} & \gate{\iota\mathrm{I}} & \gate{\mathrm{X}} & \gate{\mathrm{H}} &  \qw \\
\lstick{\ket{1}} & \gate{\mathrm{H}} &  \gate{\mathrm{X}} & \qw & \ctrl{3} & \qw  & \gate{\mathrm{X}} & \gate{\mathrm{H}} &  \qw \\
\lstick{\vdots} & \vdots & \vdots  & \vdots &  & \vdots \\
\lstick{\ket{m-2}} & \gate{\mathrm{H}} &  \gate{\mathrm{X}} & \qw & \ctrl{1} & \qw  & \gate{\mathrm{X}} &  \gate{\mathrm{H}} & \qw \\
\lstick{\ket{m-1}} & \gate{\mathrm{H}} &  \gate{\mathrm{X}} & \gate{\mathrm{H}} & \targ & \gate{\mathrm{H}} & \gate{\mathrm{X}} &  \gate{\mathrm{H}} & \qw
}
\]

\caption{Decomposition of the Grover operator.}
\label{fig:Grover_circuit}
\end{figure}

The decomposition of the Grover operator $C$ has been discussed in \cite{Batty_Decomp_Grover,Lavor_Decompose_Grover} (see Figure~\ref{fig:Grover_circuit}), which is;
\begin{equation*}
C = H^{\otimes m}X^{\otimes m} H\otimes I^{\otimes m-1} C^{(m-1)}_X([0,1,\dots,m-2],m-1)  H\otimes I^{\otimes m-1} X^{\otimes m}  H^{\otimes m},
\end{equation*}
where, $C^{(m-1)}_X(c,t)$ is the generalized Toffoli gate with a set of $m-1$ control qubits, denoted by $c$, and a target qubit $t$, i.e., it flips the value of the target only if each qubit in $c$ is at state $\ket{1}$.

\subsubsection{Decomposition of the Shift operator}
The shift operator $S$ can be expressed as
\begin{equation}
        S = \sum_{k=1}^{d}S_{k-1},\mbox{ where }S_{k-1} = \sum_{i=0}^{2^n-1}\ket{\alpha_{k-1}}\ket{v_i\oplus \Omega(k)}\bra{\alpha_{k-1}}\bra{v_i}.
\end{equation}
The operator $S_{k-1}$ shifts the walker only along the edge $\alpha_{k-1}$. Suppose $\Omega(k)$ has non-zero entries at positions $p_1,p_2,\dots,p_{l}$, then we append the extended Toffoli gates $C^{(m)}_X(c,p_j-1)$, $1\leq j\leq l$, to the quantum circuit corresponding to $S_{k-1}$, where, $c$ is the control consisting of all coin qubits and  $p_j$-th position qubit is the target. These gates do not flip the values of the target qubits unless the coin state $\ket{\alpha_{k-1}}$ is $\ket{1^m}$. Therefore, we need to first apply X gate to each coin qubit with value zero in $\ket{\alpha_{k-1}}$ and then apply the extended Toffoli gates. Thus, the sequence of gates appended to the quantum circuit corresponding to $S_{k-1}$ is given by
\begin{equation}\label{eq:S_k-1}
    \tilde{S}_{k-1} =(f(k))(C^{(m)}_X(c,p_l-1)\dots C^{(m)}_X(c,p_2-1)C^{(m)}_X(c,p_1-1)),
\end{equation}
where, $f(k)$ is an operator equivalent to appended X gates that transforms $\ket{\alpha_{k-1}}$ to $\ket{1^m}$. Thus, the sequence of operators appended to the quantum circuit corresponding to $S$ is 
\begin{equation}
    S \equiv \tilde{S}_0\tilde{S}_1\cdots \tilde{S}_{d-1}.
\end{equation}
where, $S_{k-1}$, $1\leq k\leq d$, is expressed by equation \ref{eq:S_k-1}.

\begin{Example}\label{eg:shift}
    We can understand the decomposition of the shift operator through an example of DTQW on  $Cay(\mathbb{Z}_2^4,\Omega)$, with $\Omega=\{0101,0111,1001,1010\}$. Suppose the walker is at vertex $v=1101$, then the shift operation at position state $\ket{v}=\ket{1101}$ is described as;
\begin{equation}
    \begin{split}
        S\ket{\alpha_0}\ket{1101} &= \ket{\alpha_0}\ket{1101\oplus 0101} = \ket{\alpha_0}\ket{1000} \\
        S\ket{\alpha_1}\ket{1101} &= \ket{\alpha_1}\ket{1101\oplus 0111} = \ket{\alpha_1}\ket{1010} \\
        S\ket{\alpha_2}\ket{1101} &= \ket{\alpha_2}\ket{1101\oplus 1001} = \ket{\alpha_2}\ket{0100} \\
        S\ket{\alpha_3}\ket{1101} &= \ket{\alpha_2}\ket{1101\oplus 1010} = \ket{\alpha_3}\ket{0110}.
    \end{split}
    \label{eq:shift_Q4}
\end{equation}

\begin{figure}[ht]
    \tiny
    \[
    \Qcircuit @C=1em @R=1em
    {
    \lstick{0} & \qw & \targ  &  \qw \barrier[-1em]{5}  & \qw & \targ & \qw & \qw \barrier[-1em]{5} & \qw & \targ \barrier[1em]{5} & \qw  & \qw & \qw & \qw & \qw \\
    \lstick{1} & \qw & \qw  & \qw & \qw & \qw & \targ & \qw & \qw & \qw & \qw & \qw & \targ & \qw & \qw \\
    \lstick{2} & \qw & \qw & \targ & \qw & \qw & \qw & \targ & \qw & \qw & \qw & \qw & \qw & \qw  & \qw \\
    \lstick{3} & \qw & \qw & \qw & \qw & \qw & \qw & \qw & \qw & \qw & \targ & \qw & \qw & \targ & \qw \\
    \lstick{4} & \gate{\mathrm{X}} & \ctrl{-4} & \ctrl{-2} & \gate{\mathrm{X}} & \ctrl{-4} & \ctrl{-3} & \ctrl{-2} & \gate{\mathrm{X}} & \ctrl{-4} & \ctrl{-1} & \gate{\mathrm{X}} & \ctrl{-3} & \ctrl{-1} & \qw \\
    \lstick{5} & \gate{\mathrm{X}} & \ctrl{-1} & \ctrl{-1} & \qw & \ctrl{-1} & \ctrl{-1} & \ctrl{-1} & \gate{\mathrm{X}} & \ctrl{-1} & \ctrl{-1} & \qw & \ctrl{-1} & \ctrl{-1} & \qw \\
    }
    \]
    \caption{The quantum circuit for the shift operator used in DTQW on $Cay(\mathbb{Z}_2^4,\{0101,0111,1001,1010\})$.}
    \label{fig:Q4_shift}
\end{figure}
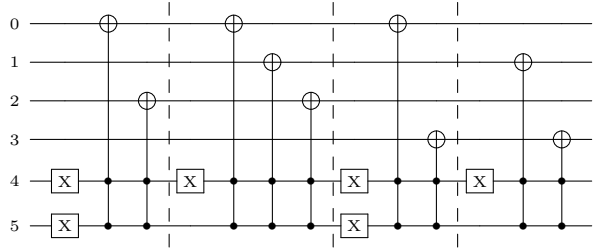

The shift operator, mentioned in Equation~\ref{eq:shift_Q4}, can be decomposed as shown in Figure~\ref{fig:Q4_shift}. For each edge $\alpha_{k-1}$, $1\leq k \leq 4$, we transform $\ket{\alpha_{k-1}}$ to $\ket{\alpha_3}\equiv \ket{11}$ by applying NOT gates, and then apply Toffoli gates  to flip the value of position qubits corresponding to non-zero entries in $\Omega(k)$. See Table~\ref{tab:trail}, where we have illustrated the transformation of coin qubits. The shift along each direction is given by;
\begin{equation}
    \begin{split}
        S_0 &= (X\otimes X) (C^{(2)}_X([4,5],0)C_X^{(2)}([4,5],2)) \\
        S_1 &= (I\otimes X) (C_X^{(2)}([4,5],0)C_X^{(2)}([4,5],1)C_X^{(2)}([4,5],2)) \\
        S_2 &= (X\otimes X) (C_X^{(2)}([4,5],0)C_X^{(2)}([4,5],3)) \\
        S_3 &= (I\otimes X) (C_X^{(2)}([4,5],1)C_X^{(2)}([4,5],3))
    \end{split}
\end{equation}
\begin{table}[t]
\centering
\begin{tabular}{|c|c|c|c|c|}
	\hline
	& $X\otimes X$ & $I\otimes X$ & $X\otimes X$ & $I\otimes X$ \\
	\hline
	00 & \textbf{11} & 10 & 01 & 00 \\
	01 & 10 & \textbf{11} & 00 & 01 \\
	10 & 01 & 00 & \textbf{11} & 10 \\
	11 & 00 & 01 & 10 & \textbf{11} \\
	\hline
\end{tabular}
\caption{Transforming each coin state $\ket{\alpha_{k-1}}$, $1\leq k\leq 4$, to $\ket{11}$ by applying NOT gates.}
\label{tab:trail}
\end{table}

\end{Example}

We now discuss how we keep track of changes made to each coin state while implementing the shift operation. For $y\in\mathbb{Z}_2^m$, define permutations $P_y$ by
\begin{equation*}
    P_y(x) = y\oplus x,\; \forall x\in\mathbb{Z}_2^m.
\end{equation*}
We construct a sequence of binary strings $B(\alpha_{k-1}) = 0^{m-r_k}1^{r_k}$, $1\leq k\leq 2^m$, where $m-r_k$ is the position of the last non-zero bit of the $k$-th binary string $\alpha_{k-1}$ in $\mathbb{Z}_2^m$. Note that $B(0^m) = 1^m$. 

Claim:
\begin{equation}\label{eq:permutation}
    P_{B(\alpha_{k-1})}(P_{B(\alpha_{k-2})}(\cdots (P_{B(\alpha_0)}(\alpha_{k-1}))\cdots)) = 1^m.
\end{equation}
% Moreover, the $2^m\times 2^m$ matrix $M$ with entries
% \begin{equation}\label{eq:symmetrix_table}
%     M(i,j) = P_{B(j)}(P_{B(j-1)}(\cdots(P_{B(1)}(e_{i-1}))));\; 1\leq i,j\leq m
% \end{equation}
% is a symmetric matrix in which each column and each row consist of all binary strings of length $m$, and $M(k,k)=1^m$. See Table~\ref{tab:trail}, in which the last $4\times 4$ submatrix is symmetric with $M(k,k)=11$, $1\leq k\leq 4$. 
If the claim is true, then $f(k)=I^{m-r_k}X^{r_k}$, $1\leq k\leq 2^m$,  transforms each coin state $\ket{\alpha_{k-1}}$ to $\ket{1^m}$, one at a time and therefore the operator given by Equation \ref{eq:S_k-1} is correct.

We now prove the claim by induction on $k$. For $k=1$, we have $\alpha_0=0^m$ and $B(\alpha_0)=1^m$. Thus, \[P_{B(\alpha_0)}(\alpha_0)=1^m\oplus 0^m=1^m.\] 
Assume that at $k$-th iteration, $\alpha_{k-1}$ gets transformed to $1^m$. Since $\alpha_{k-1}$ and $\alpha_k$ differ at the last $r$ positions, for some $1\leq r\leq m$, $\alpha_k$ gets transformed to $1^{m-r}0^r$ at $k$-th iteration. Therefore, $B(\alpha_k)=0^{m-r}1^r$ and at $(k+1)$-th iteration we get $B(\alpha_k)\oplus \alpha_k=1^m$. Hence, the claim is True.

Notice that, the transformation of a coin state corresponding to an edge alters other coin states corresponding to other edges.  At the end of the shift operation in Example~\ref{eg:shift} we retrieve the original generic coin state as shown in the last column of Table~\ref{tab:trail}. Indeed, after the $2^m$-th iterations, notice that the $i$-bit of $\alpha_{k-1}$, $1\leq i\leq m$ and $1\leq k \leq 2^m$, flips only if $B(\alpha_{j-1})$, $1\leq j\leq 2^m$, is of the form $0^{m-s}1^s$, with $s\geq i$, which corresponds to a binary string $e_{j-1}$ of the form $x_{m-1}\cdots x_i0^i$. Since the number of such binary strings are $2^{m-i}$, the $i$-th bit of $\alpha_{k-1}$ flips even number of times. Therefore, each coin state $\ket{\alpha_{k-1}}$, $1\leq k\leq 2^m$, remains unchanged after the $2^m$-th iterations.  Therefore, the shift operation is performed successfully without altering the coin state.

\subsubsection{Analysis of the quantum circuit for the shift operator} The number of $X$ gates $Num(X)$ used in the decomposition of the shift operator is equal to \begin{equation}
    Num(X) = \sum_{x\in\mathbb{Z}_2^m}wt(B(x)) = 2^{m+1}-2,
\end{equation}
where $wt(B(x))$ is the Hamming weight of $B(x)$. This can be proved by induction on $m$. For $m=1$, $Num(X)=2$ because $B(0)=1$ and $B(1)=1$. Assume by way of induction that, for $m=k$, $Num(X)=2^{k+1}-1$. Then, for $m=k+1$, notice that if $x\in\mathbb{Z}_2^m$ then $0x,1x\in\mathbb{Z}_2^{m+1}$ and,
\begin{equation}
    B(bx)=
    \begin{cases}
        B(x),& \mbox{if }x\neq 0^m \\
        B(x)+1, & \mbox{if }x=0^m.
    \end{cases}
\end{equation} 
where $b=0$ or $1$. Therefore, $Num(X)$ for $m=k+1$ is equal to
\[
2\times(2^{k+1}-2)+2 = 2^{k+2}-2.
\]
$Num(X)$ is of order $\mathcal{O}(|\Omega|)$. The number of generalized Toffoli gates used is equal to the sum of Hamming weights of all elements of the generating set $\Omega$, which is 
\begin{equation}
    Num(C^{(m)}_X) = \sum_{x\in\Omega}wt(x),\qquad wt(x)=\mbox{number of 1's in }x.
\end{equation}

\subsection{Quantum circuit for cubelike graph of arbitrary degree}
\label{subsec: gen_cubelike_circuit}
If the degree of the cubelike graph is not a power of $2$ (for example $Q_3$)  then we have to make certain modifications to implement our circuit. Consider a cubelike graph $Cay(\mathbb{Z}_2^n,\Omega)$, with $\Delta=|\Omega|$. Let $m$ be a positive integer satisfying $2^{m-1}<\Delta\leq 2^m$. The Grover coin $C=2\ket{D}\bra{D}-I$ is a $\Delta\times \Delta$ matrix, where $\Delta$ may not be equal to $2^m$. We, therefore, define a new $2^m\times 2^m$ operator $C'$ by
\begin{equation}\label{eq:gen_grover}
    C' = \begin{bmatrix} 2\ket{D}\bra{D} & \textbf{0} \\ \textbf{0} & \textbf{0} \end{bmatrix} - I = 2\ket{D'}\bra{D'}-I,
\end{equation}
where, $\ket{D'}=\frac{1}{\sqrt{\Delta}}\sum_{k=1}^{\Delta}\ket{\alpha_{k-1}}\in\mathbb{C}^{2^m}$ is a projection of the diagonal state $\ket{D}$ in the higher dimensional Hilbert space $\mathcal{H}_{C'}=\mathbb{C}^{2^m}$ that contains the coin space $\mathcal{H}_C=\mathbb{C}^{\Delta}$ as a subspace. In other words, $\mathcal{H}_{C'}$ is a new coin space with the computational basis $\{\ket{\alpha_{k-1}}:1\leq k \leq 2^m\}$, and the quantum walk occurs in $Cay(\mathbb{Z}_2^n,\Omega')$, where $\Omega'$ has $2^m$ elements containing $\Omega$, i.e., $Cay(\mathbb{Z}_2^n,\Omega)$ is a subgraph of $Cay(\mathbb{Z}_2^n,\Omega')$. The new coin operator $C'$ changes the coefficients of the initialized vector, which is $\ket{D'}$, and therefore, the coefficents of coin states $\ket{\alpha_{k-1}}$, with $k\geq d+1$, remain $0$ throughout the evolution, i.e. the generic coin state $\ket{c}$, in our case, is
\begin{equation}
    \ket{c} = \sum_{k=1}^{\Delta}\lambda_k\ket{\alpha_{k-1}} + \sum_{k=\Delta+1}^{2^m}\ket{\alpha_{k-1}}.
\end{equation}
We define a new shift operator $S'$ by
\begin{equation}
    S' = \sum_{k=1}^{\Delta}\sum_{a=0}^{2^n-1}\ket{\alpha_{k-1}}\ket{v_a\oplus \Omega(k)}\bra{\alpha_{k-1}}\bra{v_a} + \sum_{k=\Delta+1}^{2^m}\sum_{a=0}^{2^n-1}\ket{\alpha_{k-1}}\ket{v_a}\bra{\alpha_{k-1}}\bra{v_a}.
\end{equation}
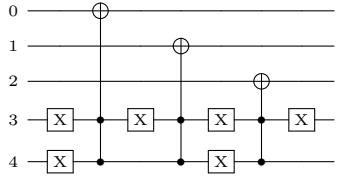
\begin{figure}[h]
    \tiny
    \centering
    \[
    \Qcircuit @C=1em @R=1em
    {
    \lstick{0} & \qw & \targ & \qw & \qw & \qw & \qw & \qw & \qw \\
    \lstick{1} & \qw & \qw & \qw & \targ & \qw & \qw & \qw & \qw \\
    \lstick{2} & \qw & \qw & \qw & \qw & \qw & \targ & \qw & \qw \\
    \lstick{3} & \gate{\mathrm{X}} & \ctrl{-3} & \gate{\mathrm{X}} & \ctrl{-2} & \gate{\mathrm{X}} & \ctrl{-1} & \gate{\mathrm{X}} & \qw \\
    \lstick{4} & \gate{\mathrm{X}} & \ctrl{-1} & \qw & \ctrl{-1} & \gate{\mathrm{X}} & \ctrl{-1} & \qw & \qw 
    }
    \]
    \caption{Quantum circuit for $Q_3$}
    \label{fig:Q3_shift}
\end{figure}

The new shift operator $S'$ fixes $\ket{\alpha_{k-1}}\ket{v_a}$, $0\leq a\leq 2^n-1$, if $k\geq \Delta+1$, i.e., $S'\ket{\alpha_{k-1}}\ket{v_a}=\ket{\alpha_{k-1}}\ket{v_a}$. The Figure~\ref{fig:Q3_shift} represents the quantum circuit for $\mathcal{Q}_3$. Notice that the circuit does not shift a position state along the direction $\ket{\alpha_3}\equiv\ket{11}$. Finally, the original coin state is retrieved at the end of the shift operation because each quantum wire contains even number of X gates.

%%%%%%%%%%%%%%%%%%%%%%%%%%%%%%%%%%%%%%%%%%%%%%%%%%%%%%%%%%%%%%%%%%%%%%%%%%%%%%%%%%%%%%%%%%%
%%%%%%%%%%%%%%%%%%%%%%%%%%%%%%%%%%%%%%%%%%%%%%%%%%%%%%%%%%%%%%%%%%%%%%%%%%%%%%%%%%%%%%%%%%%

\section{Application to hitting times on cubelike graphs}
\label{sec:Results}
We use Qiskit \cite{Qiskit} to implement DTQW on IBM's Quantum simulators and quantum computers. The Qiskit simulator \textit{qasm\_simulator} runs locally and other IBM's simulators such as \textit{simulator\_mps}, \textit{simulator\_extended\_stabilizer}, etc, are accessed via IBM provider \textit{ibm-q}.

\subsection{DTQW on Quantum computers} 
\begin{figure}[h]
    \centering
    \begin{subfigure}{.35\textwidth}
      \includegraphics[width=\textwidth]{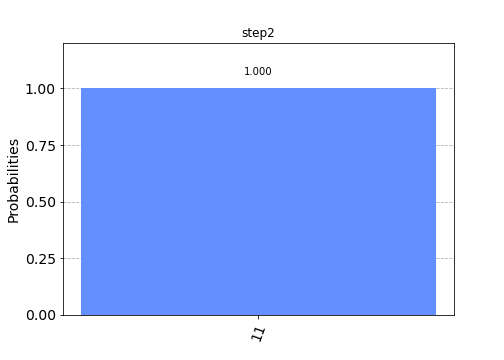}
      \caption{\label{fig:qasm_Q2}}
    \end{subfigure}
    \begin{subfigure}{.35\textwidth}
      \includegraphics[width=\textwidth]{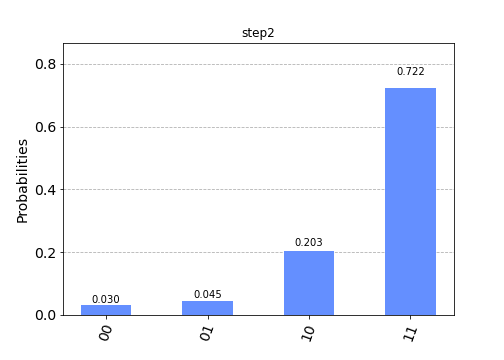}
      \caption{\label{fig:ibmq_Q2}}
    \end{subfigure}
    \caption{Probability distribution of DTQW on $\mathcal{Q}_2$ after two steps when run on Qiskit simulator (\subref{fig:qasm_Q2}) \textit{qasm\_simulator} and on IBM's quantum computer (\subref{fig:ibmq_Q2}) \textit{ibmq\_manila}.}
    \label{fig:real_vs_simulator_Q2}
\end{figure}

The quantum circuit for DTWQ on $\mathcal{Q}_2$, see Figure~\ref{fig:Q2_circuit}, was run on IBM's quantum computer \textit{ibmq\_manila v1.0.3}, the result of which is compared with the result on \textit{qasm\_simulator}, as shown in Figure~\ref{fig:real_vs_simulator_Q2}. The quantum circuits corresponding to higher dimensional cubelike graphs could not be run successfully on real quantum computers due to unavoidable gate errors and noise from the environment around the quantum computer. In case of $\mathcal{Q}_3$, the errors and noise were low and therefore could be run successfully.

\subsection{Hitting times of Hypercubes and Augmented cubes}
Upon implementing quantum circuits for DTQW on cubelike graphs, we have observed that the specific vertex, called the target vertex, to which the walker reach with highest probability, is unique and equal to the binary XOR operations of elements of the generating set $\Omega$, i.e.,
\begin{equation}
    \mbox{target vertex }=\bigoplus_{x\in\Omega}x.
\end{equation}

\begin{figure}[h]
    \centering
    \begin{subfigure}[]{\textwidth}
    \centering
        \includegraphics[scale=.2]{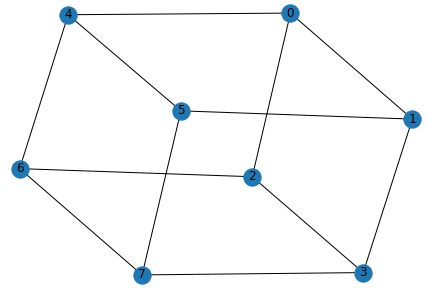}
        \includegraphics[width=9cm,height=4cm]{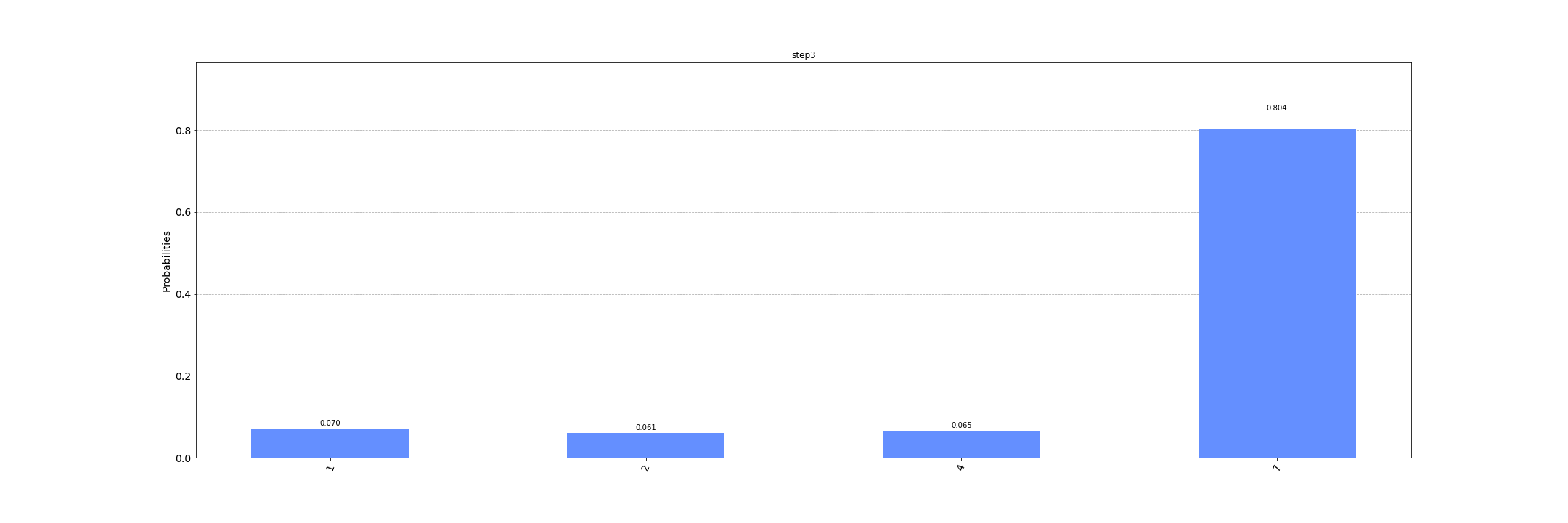}
        \caption{\label{fig:Q3_DTQW} DTQW on $\mathcal{Q}_3$ with steps $T=3$, target vertex $7\equiv 111$, and target probability $0.804$.}
    \end{subfigure}    
    \begin{subfigure}{\textwidth}
        \includegraphics[scale=.2]{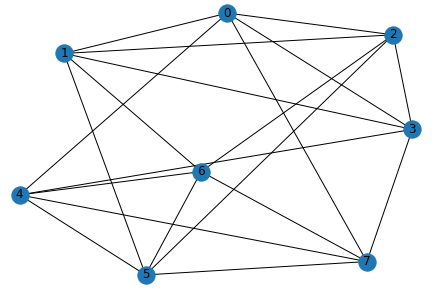}
        \includegraphics[width=9cm,height=4cm]{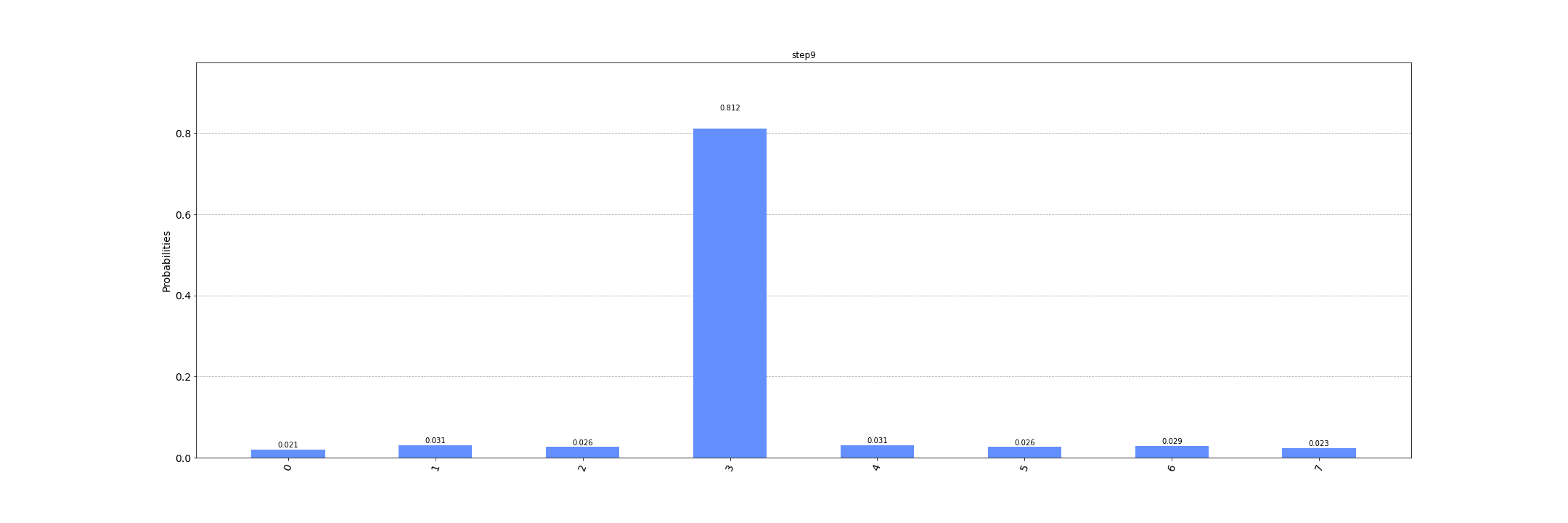}
        \caption{\label{fig:AQ3_DTQW}DTQW on $\mathcal{AQ}_3$ with steps $T=9$, target vertex $3\equiv 011$, and target probability $0.812$.}
    \end{subfigure} 
    \caption{Probability distribution of Hypercubes and Augmented cubes of dimension $3$. Target vertices correspond to the tallest bar in each bar graph. }
    \label{fig:dim_3}
\end{figure}

\begin{figure}[h!]
    \centering
    \begin{subfigure}[]{\textwidth}
    \centering
        \includegraphics[scale=.2]{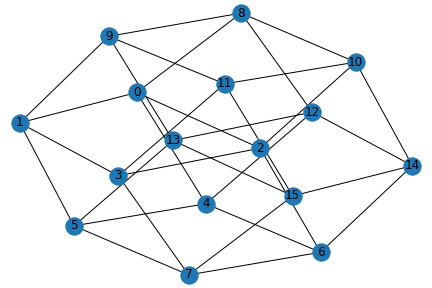}
        \includegraphics[width=9cm,height=4cm]{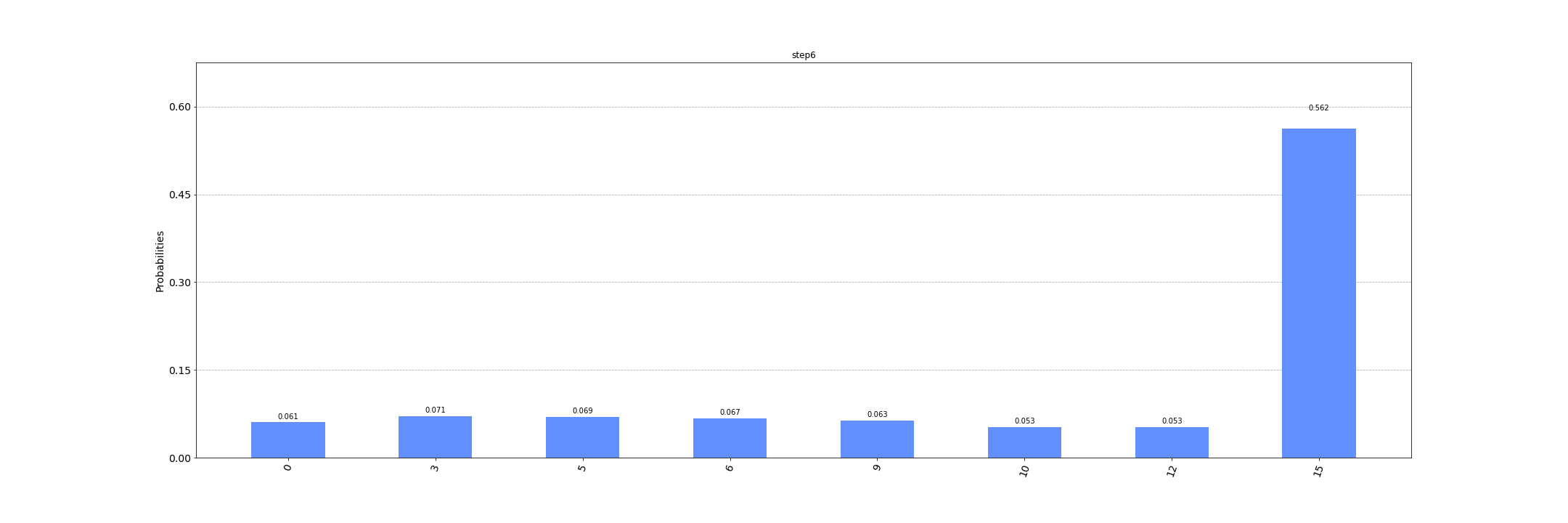}
        \caption{\label{fig:Q4_DTQW} DTQW on $\mathcal{Q}_4$ with steps $T=6$, target vertex $15\equiv 1111$, and target probability $0.562$.}
    \end{subfigure}    
    \begin{subfigure}{\textwidth}
        \includegraphics[scale=.2]{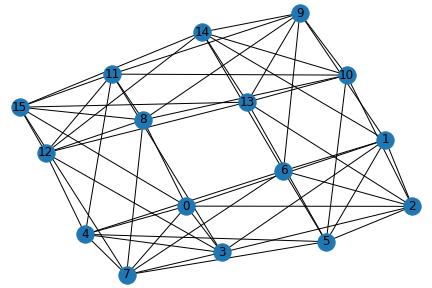}
        \includegraphics[width=9cm,height=4cm]{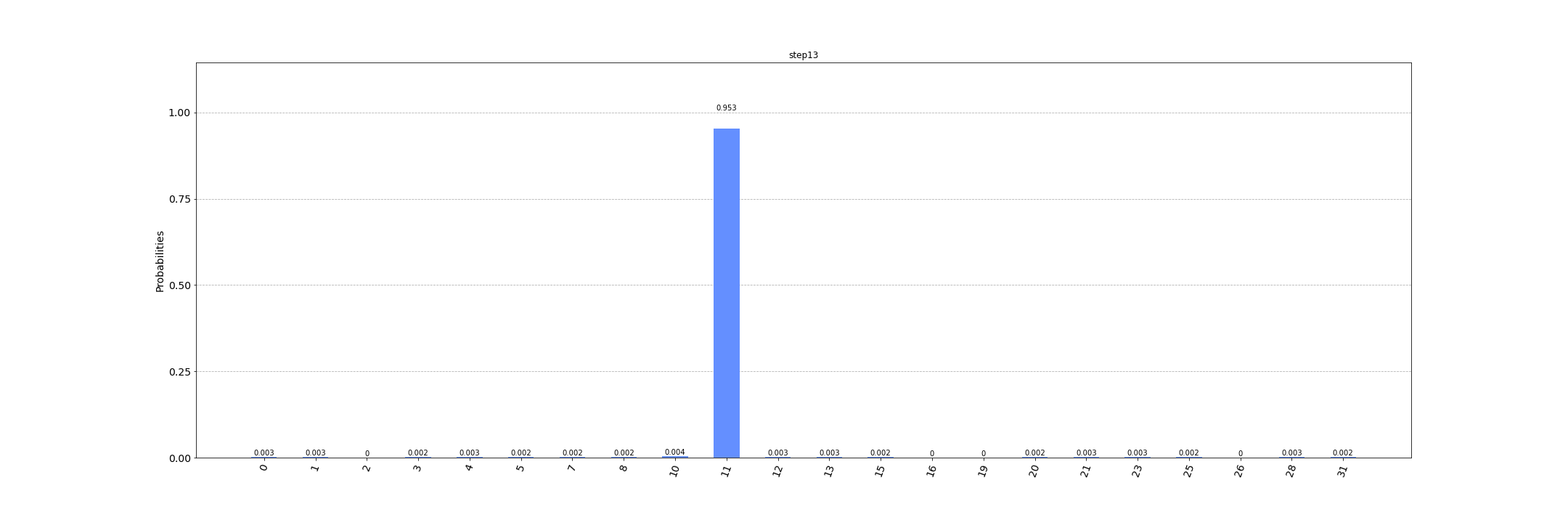}
        \caption{\label{fig:AQ4_DTQW}DTQW on $\mathcal{AQ}_4$ with steps $T=11$, target vertex $4\equiv 0100$, and target probability $0.993$.}
    \end{subfigure} 
    \begin{subfigure}[]{\textwidth}
    \centering
        \includegraphics[scale=.2]{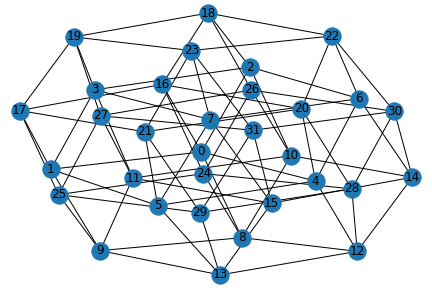}
        \includegraphics[width=9cm,height=4cm]{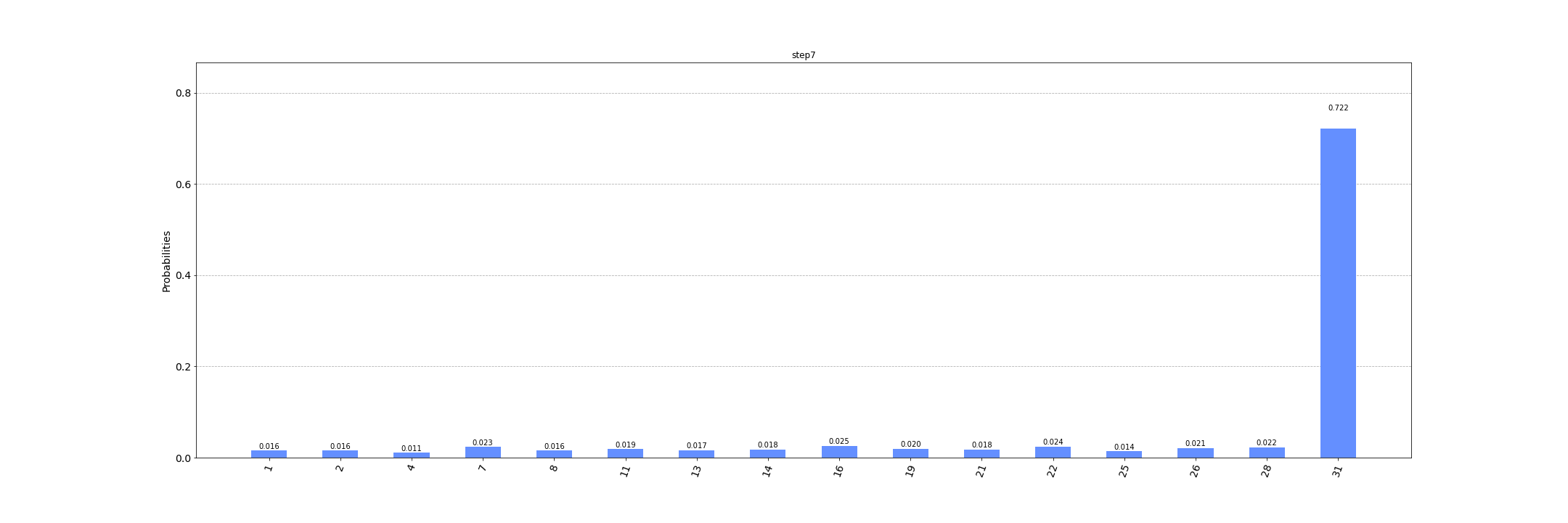}
        \caption{\label{fig:Q5_DTQW} DTQW on $\mathcal{Q}_5$ with steps $T=7$, target vertex $31\equiv 11111$, and target probability $0.722$.}
    \end{subfigure}    
    \begin{subfigure}{\textwidth}
        \includegraphics[scale=.2]{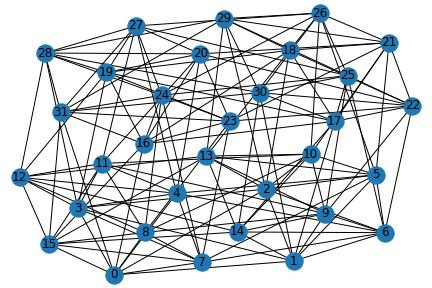}
        \includegraphics[width=9cm,height=4cm]{Figures/Augmented_cubes/cube_5_deg_9_step_13.png}
        \caption{\label{fig:AQ5_DTQW}DTQW on $\mathcal{AQ}_5$ with steps $T=13$, target vertex $11\equiv 01011$, and target probability $0.953$.}
    \end{subfigure} 
    \caption{Probability distribution of Hypercubes and Augmented cubes of dimension $4$ and $5$.}
    \label{fig:dim_4&5}
\end{figure}

\clearpage
\begin{table}[]
\small
    \centering
    \begin{minipage}{.45\textwidth}
    \setcounter{table}{0}
    \renewcommand{\thetable}{\Alph{table}}
    \centering
        \begin{tabular}[5pt]{|c|c|c|c|}
            \hline
            $n=\Delta$ & T & Target & $p$   \\
            \hline
            3&3&$ 111$&0.804\\
            \hline
            4&6&$ 1111$&0.562\\
            \hline
            5&7&$ 11111$&0.722\\
            \hline
            6&10&$ 1^6$&0.816 \\
            \hline
            7&11&$ 1^7$&0.912 \\
            \hline
            8&12&$ 1^8$&0.954 \\
            \hline
            9&13&$ 1^9$&0.950 \\
            \hline
            10&14&$1^{10}$&0.901 \\
            \hline
            11&17&$1^{11}$&0.927 \\
            \hline
            12&18& $ 1^{12}$ &0.956 \\
            \hline
            13&19&$1^{13}$&0.947 \\
            \hline
            14&22&$ 1^{14}$&0.929 \\
            \hline
            15&23&$ 1^{15}$&0.961 \\
            \hline
            16&24&$ 1^{16}$&0.960 \\
            \hline
        \end{tabular}
        \caption{\label{tab:Qn} $\mathcal{Q}_n$.}
    \end{minipage}  
    \begin{minipage}{.45\textwidth}
    \centering
    \setcounter{table}{1}
    \renewcommand{\thetable}{\Alph{table}}
        \begin{tabular}[2pt]{|c|c|c|c|c|}
            \hline
            n & $\Delta$ & T & Target & $p$   \\
            \hline
            3&5&9&$ 011$&0.812\\
            \hline
            4&7&11&$ 0100$&0.993\\
            \hline
            5&9&13&$ 01011$&0.953\\
            \hline
            6&11&17&$ (01)^200$&0.928 \\
            \hline
            7&13&19&$  (01)^31$&0.919 \\
            \hline
            8&15&23&$ (01)^300$&0.969 \\
            \hline
            9&17&25&$ (01)^41$&0.926 \\
            \hline
            10&19&29&$(01)^400$&0.955 \\
            \hline
            11&21&31&$(01)^51$&0.919 \\
            \hline
            12&23&35&$(01)^500$&0.954 \\
            \hline
            13&25&39&$(01)^61$&0.958 \\
            \hline
            14&27&41&$(01)^600$&0.962 \\
            \hline
            15&29&45&$(01)^71$&0.977 \\
            \hline
            16&31&47&$(01)^700$&0.961 \\
            \hline
        \end{tabular}
        \caption{\label{tab:AQn} $\mathcal{AQ}_n$.}
    \end{minipage}
    \caption*{In Table~\ref{tab:Qn} and Table~\ref{tab:AQn}, $n$ denotes dimension  of $\mathcal{Q}_n$ and $\mathcal{AQ}_n$. $T$ is the hitting time with probability $p$, and $\Delta$ is the degree.}
    \label{tab:Qn+AQn}
\end{table}

\begin{figure}[htp]
    \centering
    \begin{subfigure}{.45\textwidth}
      \includegraphics[width=\textwidth]{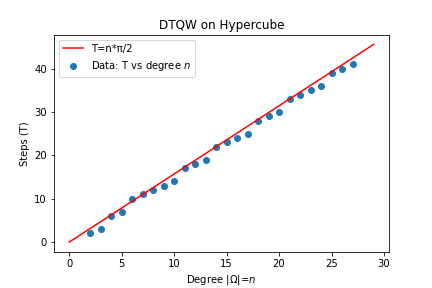}
      \caption{\label{fig:T_vs_deg_n+0}}
    \end{subfigure}
    \begin{subfigure}{.45\textwidth}
      \includegraphics[width=\textwidth]{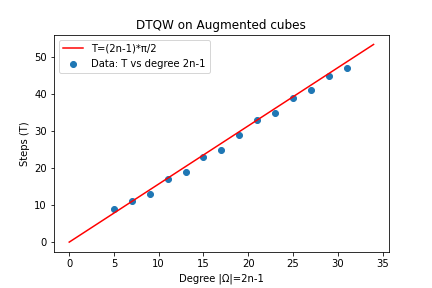}
      \caption{\label{fig:T_vs_deg_2n-1}}
    \end{subfigure}
    \caption{Plot of the hitting time $T$ vs the degree (\subref{fig:T_vs_deg_n+0}) $|\Omega|=n$ of the Hypercube and (\subref{fig:T_vs_deg_2n-1}) $|\Omega|=2n-1$ of the Augmented cube.}
    \label{fig:T_vs_Hyper+Augmented}
\end{figure}

In Figure~\ref{fig:dim_3} and ~\ref{fig:dim_4&5} the target probabilities for Augmented cubes are higher than that of Hypercubes of same dimensions. In either case, the target probability gets closer to $1$ as the dimension increases. In Table~\ref{tab:Qn} and \ref{tab:AQn}, we have displayed the hitting time $T$ along with target and target probability corresponding to dimensions of $\mathcal{Q}_n$ and $\mathcal{AQ}_n$. In case of Hypercubes, $T$ is linear with $n$ that tallies with the theoretical result stated by Kempe in \cite{Kempe2005}, while in Augmented cubes it is linear with the degree of the graph. See Figure~\ref{fig:T_vs_deg_2n-1}, where we have shown that $T$ is linear with the degree $2n-1$ of $\mathcal{AQ}_n$. 

\begin{figure}[htp]
    \centering
    \begin{subfigure}{.45\textwidth}
      \includegraphics[width=\textwidth]{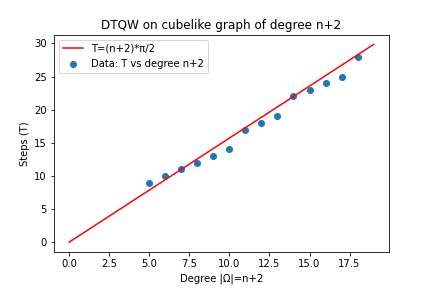}
      \caption{\label{fig:T_vs_deg_n+2}}
    \end{subfigure}
    \begin{subfigure}{.45\textwidth}
      \includegraphics[width=\textwidth]{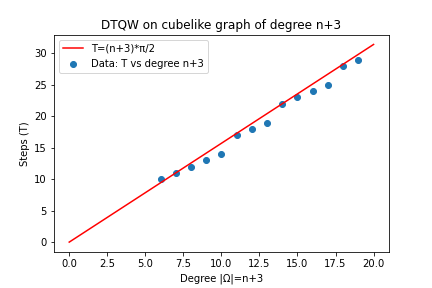}
      \caption{\label{fig:T_vs_deg_n+3}}
    \end{subfigure}
    \begin{subfigure}{.45\textwidth}
      \includegraphics[width=\textwidth]{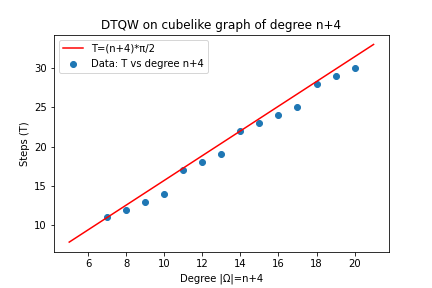}
      \caption{\label{fig:T_vs_deg_n+4}}
    \end{subfigure}
    \begin{subfigure}{.45\textwidth}
      \includegraphics[width=\textwidth]{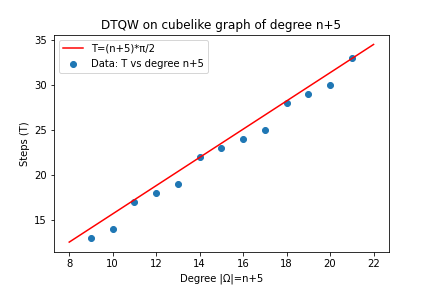}
      \caption{\label{fig:T_vs_deg_n+5}}
    \end{subfigure}
    \begin{subfigure}{.45\textwidth}
      \includegraphics[width=\textwidth]{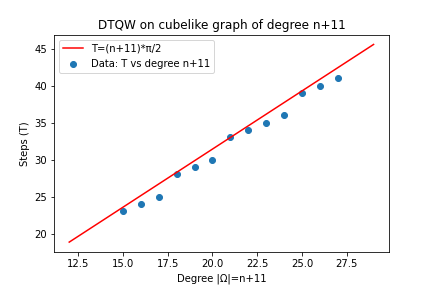}
      \caption{\label{fig:T_vs_deg_n+11}}
    \end{subfigure}
    \begin{subfigure}{.45\textwidth}
      \includegraphics[width=\textwidth]{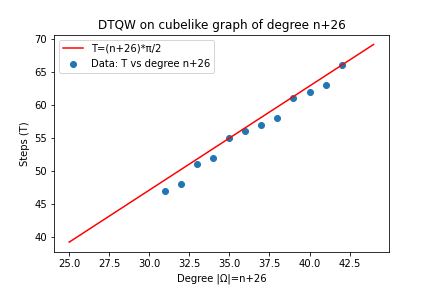}
      \caption{\label{fig:T_vs_deg_n+26}}
    \end{subfigure}
    \caption{Plots of steps T required to attain the target probability verses the degree $|\Omega|$ of cubelike graphs of degree $n+k$, where $k$ is fixed and $n$ is the dimension.}
    \label{fig:T_vs_degree}
\end{figure}

\subsection{A conjecture on hitting times of general cubelike graphs}
\begin{figure}[htp]
    \centering
    \begin{subfigure}{.45\textwidth}
      \includegraphics[width=\textwidth]{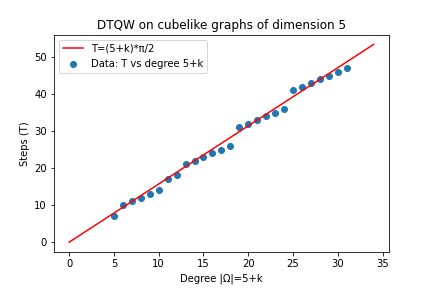}
      \caption{\label{fig:T_vs_dim_5}}
    \end{subfigure}
    \begin{subfigure}{.45\textwidth}
      \includegraphics[width=\textwidth]{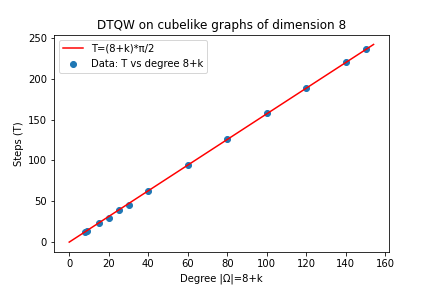}
      \caption{\label{fig:T_vs_dim_8}}
    \end{subfigure}
    \begin{subfigure}{.45\textwidth}
      \includegraphics[width=\textwidth]{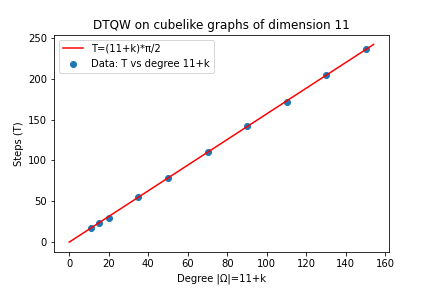}
      \caption{\label{fig:T_vs_dim_11}}
    \end{subfigure}
    \begin{subfigure}{.45\textwidth}
      \includegraphics[width=\textwidth]{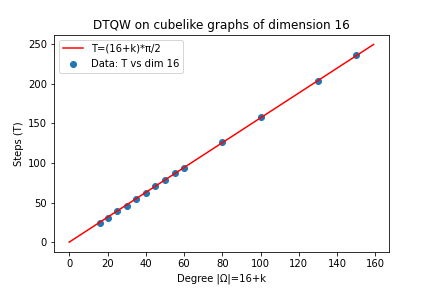}
      \caption{\label{fig:T_vs_dim_16}}
    \end{subfigure}
    \caption{Plots of steps T requried to attain the target probability verses the degree $|\Omega|=n+k$ of cubelike graphs of fixed dimension n.}
    \label{fig:T_vs_fixed_dim}
\end{figure}

The result mentioned by J. Kempe in \cite{Kempe2005} can be generalized to other cubelike graphs. In Figure~\ref{fig:T_vs_degree} and \ref{fig:T_vs_fixed_dim} we have shown plots of steps T verses degrees $n+k$ of some cubelike graphs, where $T$ is the number of iterations required to hit the target vertex, and $n$ is the dimension of cubelike graph of degree $n+k$. In the first figure, we fix the value of $k$ and increase $n$, and in the second figure, we fix the dimension $n$ and increase the value of $k$. In either case, we find that $T$ verses $n+k$ plot is linear and follows the relation;
\begin{equation}
    T \approx (n+k)\times \pi/2.
\end{equation}
It is to be noted that it is just a coincidence that $T$ vs dimension plot is same as $T$ vs degree plot for DTQW on Hypercubes.

% \begin{figure}[htp]
%     \centering
%     \includegraphics[scale=.4]{T_vs_n+k_plot/T_vs_degree.png}
%     \caption{Plot of T verses degrees of several cubelike graphs.}
%     \label{fig:T_vs_degree}
% \end{figure}

The observations and remarks made above lead to the following conjecture.
\begin{conjecture}
Let $\Gamma=Cay(\mathbb{Z}_2^n,\Omega)$ be an $n$-dimensional Cubelike graph with degree $\Delta=|\Omega|$. Define the target vertex by;
\begin{equation}
    v_{targ} = \bigoplus_{x\in\Omega} x.
\end{equation}
Then, there exists a hitting time $T\approx\frac{\pi \Delta}{2}$ such that the target probability $p_{targ}$, the probability by which DTWQ reaches $v_{targ}$, is asymptotically equal to 1. Moreover, the parity of $T$ is same as that of $\Delta$, i.e., $T$ is even only if $\Delta$ is even. 
\end{conjecture}

%%%%%%%%%%%%%%%%%%%%%%%%%%%%%%%%%%%%%%%%%%%%%%%%%%%%%%%%%%%%%%%%%%%%
%%%%%%%%%%%%%%%%%%%%%%%%%%%%%%%%%%%%%%%%%%%%%%%%%%%%%%%%%%%%%%%%%%%%
%%%%%%%%%%%%%%%%%%%%%%%%%%%%%%%%%%%%%%%%%%%%%%%%%%%%%%%%%%%%%%%%%%%%

\section{Conclusion}
Quantum random walks and their hitting times are an important area of study in quantum computation due to their applications in development of faster quantum algorithms. In this paper we have implemented efficient quantum circuits for discrete quantum random walks on families of cubelike graphs such as hypercubes and augmented cubes. Our implementations show that the hitting-times of all cubelike graphs is asymptotically linear in the degree $\Delta$ of the graphs. That is, for the hitting time $\frac{\pi \Delta}{2}$, probability that the walker is found at the target vertex approaches $1$ as  $\Delta$ approaches infinity. Our circuits run on IBM's quantum computing platform Qiskit, both on real quantum computers as well as simulators. 

\begin{figure}[htp]
    \centering
    \includegraphics[scale=.4]{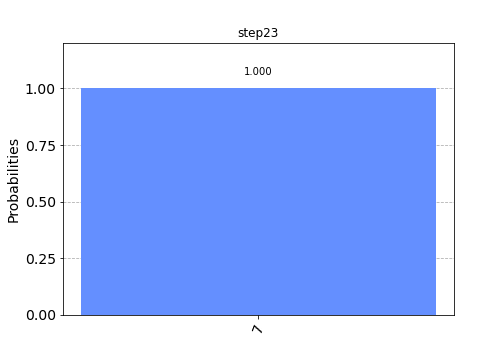}
    \caption{Probability distribution of $\mathcal{Q}_3$ after step $T=23$.}
    \label{fig:Q3_step23}
\end{figure}

\begin{figure}[htp]
    \centering
    \includegraphics[scale=.4]{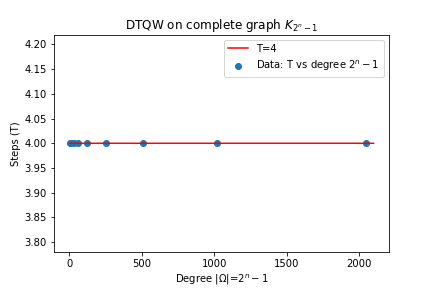}
    \caption{Plot for hitting time $T$ verses degree $2^n-1$ of complete graph on $2^n$ vertices.}
    \label{fig:T_vs_K_n}
\end{figure}

We note  that $T\approx \frac{\Delta\pi}{2}$ is not necessarily  the minimum hitting time or its multiple for a $\Delta$-regular cubelike graph. For example, in DTQW on $\mathcal{Q}_3$, see Figure~\ref{fig:Q3_step23}, the walker hits the target vertex $\ket{111}$ after steps $T=23$ with probability 1. Another example is that of a complete graph. The cubelike graphs $Cay(\mathbb{Z}_2^n,\Omega)$, with $\Omega=\mathbb{Z}_2^n\backslash\{0^n\}$, are complete graphs. In Figure~\ref{fig:T_vs_K_n}, the hitting time for the complete graphs on $2^n$ vertices has been shown constant with the value $T=4$, with target probability equal to 1, irrespective of dimensions and degrees of the graphs. We have theoretically verified that $T=4$ is the minimum hitting time with the target probability equal to 1 for the complete graphs on $2^n$ vertices. It will of interest to find other hitting times for a cubelike graph, specially the minimum one for which the walker hit the target with high probability or with probability equal to $1$.

Further, it would be interesting to study the same problem on other types of regular graphs particularly cayley graphs that are not cubelike graphs. Another implementation problem to study along the same lines is of perfect state transfer on regular graphs in the continuous time random walks. It would be interesting to build efficient quantum circuits for the perfect state transfer problem.

%%%%%%%%%%%%%%%%%%%%%%%%%%%%%%%%%%%%%%%%%%%%%%%%%%%%%%%%%%%%%%%%%%%%%%%%%%%%%%%%%%%%%%%%%%%
%%%%%%%%%%%%%%%%%%%%%%%%%%%%%%%%%%%%%%%%%%%%%%%%%%%%%%%%%%%%%%%%%%%%%%%%%%%%%%%%%%%%%%%%%%%

\bibliographystyle{acm}
\bibliography{QRW_Implementation_2021}

\end{document}